\documentclass[a4paper,11pt]{article}
\usepackage{pos}
\usepackage{caption}
\usepackage{subcaption}
\usepackage{wrapfig}
\usepackage{enumitem}
\title{Search for Ultra-High-Energy Neutrinos at the Pierre Auger Observatory: New Triggers, Methods, and Constraints}
 \ShortTitle{Search for Ultra-High-Energy Neutrinos at the Pierre Auger Observatory with EM triggers}

\author*[a]{Srijan Sehgal}

\affiliation[a]{Bergische Universität Wuppertal, Wuppertal, Germany}

\onbehalf{for the Pierre Auger Collaboration$^b$}
\affiliation[b]{Observatorio Pierre Auger, Av.\ San Mart{\'\i}n Norte 304, 5613 Malarg\"ue, Argentina\\
Full author list: {\rm\url{https://www.auger.org/archive/authors_icrc_2025.html}}}

\emailAdd{spokespersons@auger.org}

\abstract{The Pierre Auger Observatory has the capability to identify neutrino-induced extensive air showers above $10^{17}$\,eV by using its large Surface Detector (SD) array. Data from the Observatory have been used to set some of the most stringent upper limits to the neutrino flux in the ultra-high energy (UHE) range. The data have also been used for follow-up detection of transient events in the context of multi-messenger astrophysics. In mid-2013, two additional SD triggers (Time-over-Threshold-deconvolved (ToTd) and Multiplicity-of-Positive Steps (MoPS)) were shown to increase the detection capability for the neutrino-induced air showers in the energy regime below $10^{19}$\,eV by a factor of 5-10.
This contribution will give an overview of the ongoing work regarding the searches for UHE neutrinos at the Pierre Auger Observatory. The impact of the ToTd and MoPS triggers for neutrino search in the zenith angle range of $60^{\circ} < \theta < 75^{\circ}$ is discussed. A novel neutrino identification method, which integrates these triggers, is applied to observational data to look for neutrino-like events using a \textit{blind} search strategy. New constraints to point-like sources of UHE neutrinos will be presented for the angular range explored.}

\ConferenceLogo{Pos_ICRC2025_logo}

\FullConference{%
39th International Cosmic Ray Conference (ICRC2025)\\
 15–24 July 2025\\
Geneva, Switzerland\\}

\begin{document}
\maketitle

\section{Introduction}

Ultra-high-energy (UHE) neutrinos ($E > 10^{17}$~eV) are key to probing extreme astrophysical environments and understanding the origin of UHE cosmic rays (UHECRs)~\citep{Coleman:2022abf}. Unlike charged cosmic rays or high-energy photons, neutrinos travel vast distances unaffected by magnetic fields or significant attenuation, but their weak interactions make direct detection at UHE challenging. An effective method involves detecting extensive air showers (EAS) induced by neutrino charged-current (CC) or neutral-current (NC) interactions in the atmosphere or Earth, a technique employed by the Pierre Auger Observatory. Located in Argentina and covering 3000~km$^2$, Pierre Auger Observatory combines a Surface Detector (SD) array of 1660 water-Cherenkov detectors (WCDs) with a Fluorescence Detector (FD) system comprising 27 fluorescence telescopes at four locations on its periphery.

The nearly continuous operation of the SD makes it well-suited for neutrino searches, having set one of the strongest limits on the diffuse UHE neutrino flux~\cite{auger_neutrino_limit}. Neutrino-induced showers are classified as downward-going (DG) or Earth-skimming (ES), with DG further split into high and low zenith angles: DGH, $\theta \in  [75^{\circ}, 90^{\circ})$ and DGL $\theta \in  [60^{\circ}, 75^{\circ}]$. This analysis focuses on the DGL range using SD data from 1$^\mathrm{st}$ January 2014 up to 31$^\mathrm{st}$ December 2021. A new selection incorporating all electromagnetic (EM) triggers is developed and tested on simulations, improving sensitivity to neutrino-induced EASs before being applied to data.

\section{The electromagnetic triggers}

The SD employs a hierarchical trigger system to identify air showers, starting from local station-level triggers up to array-wide and physics-level triggers~\citep{PierreAuger:2010zof}. Each WCD generates local triggers based on time-dependent signals, expressed in vertical equivalent muon (VEM) units, which corresponds to the total charge deposited by a single muon passing vertically through the water volume~\citep{PierreAuger:2010zof}. The initial triggers include the \textit{threshold} (TH) trigger, optimized for muonic signals, and the \textit{time-over-threshold} (ToT) trigger, more sensitive to EM components of an EAS. These are tuned to reject low-energy background, such as single muons and consequently pure EM signal.

To enhance sensitivity to EM-dominated showers, especially from neutrinos or photons, two additional triggers — \textit{time-over-threshold-deconvolved} (ToTd) and \textit{multiplicity-of-positive-steps} (MoPS) — were introduced in June 2013. ToTd compresses long exponential tail of diffusely reflected Cherenkov light associated with muon before applying a ToT condition to reduce background, while MoPS targets long non-smooth, low-amplitude signals typical of EM cascades. Both operate at a few Hz and are combined in a logical OR with ToT, effectively lowering the array's energy threshold for detecting EM-rich air showers.

\section{Search for UHE neutrinos \boldmath$60^{\circ} < \theta < 75^{\circ}$ \unboldmath}

A key challenge in identifying UHE neutrino-induced EASs is distinguishing them from those initiated by UHECRs, such as protons or nuclei. The SD-based search at the Pierre Auger Observatory focuses on “young” showers — electromagnetic-rich and developing deep in the atmosphere—at zenith angles $\theta > 60^\circ$, where they contrast with “old,” muon-dominated CR showers. Young showers produce broader, lower-peak SD signals, while old showers yield narrow, high-peak signals. At $\theta < 60^\circ$, this distinction blurs, limiting neutrino searches to more inclined geometries.

Selection relies on signal-based variables like the fraction of ToT triggers, tuned to broader signals~\citep{auger_neutrino_limit}. Newer EM-sensitive triggers, ToTd and MoPS, further enhance sensitivity to low-energy ($\leq$1~EeV) neutrinos by rejecting isolated muons. Another key variable is the Area-over-Peak (AoP), the ratio of the integrated signal to its peak, averaged over photo-multipliers; muon-like signals yield AoP values near 1, while EM signals show AoP $>$ 1.

\subsection{Impact of ToTd and MoPS triggers }
\begin{figure}[ht!]
  \centering
  \includegraphics[width=14.5cm]{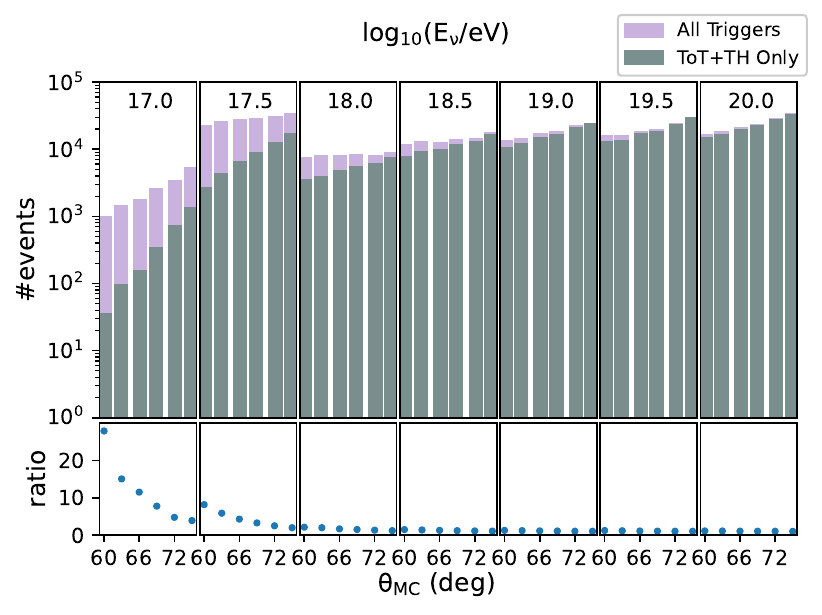}
  \caption{Reconstructed number of simulated neutrino events for all energies and channels (CC and NC) as a function of simulated zenith angle $\theta_{\rm{MC}}$ for the sample with \textit{All} (ToTd, MoPS + ToT, TH) triggers (purple bars) and only ToT+TH triggers (dark green bars). The bottom panel shows the ratio of the two samples.}
  \label{fig:Events_vs_angle_summary}
\end{figure}

To evaluate the improvement in neutrino detection sensitivity from the addition of ToTd and MoPS triggers, simulated neutrino events reconstructed without these triggers were directly compared to the same events reconstructed with all triggers enabled.
Figure~\ref{fig:Events_vs_angle_summary} presents the total number of reconstructed events with and without ToTd and MoPS triggers across all energies and interaction channels (CC+ NC) as a function of the zenith angle.

The increase in reconstructed events is most pronounced at energies below $\sim$ 1\,EeV, largely due to the enhanced sensitivity of the ToTd and MoPS triggers to weaker, more electromagnetic-dominated signals, which are characteristic of lower-energy neutrino-induced showers. This gain decreases at higher energies, where the \textit{old} ToT and TH triggers already operate with high efficiency. Additionally, the effectiveness of ToTd and MoPS triggers varies with the zenith angle. At smaller zenith angles, neutrino showers retain a stronger electromagnetic component upon reaching the ground, making them more likely to be captured by ToTd and MoPS. At larger zenith angles, ToTd and MoPS triggers improve identification primarily for neutrinos that interact closer to the detector array, where the shower still contains an electromagnetic signature. Additionally, for shallower interaction points, the resulting showers become increasingly muonic due to atmospheric attenuation, reducing the added benefit of the ToTd and MoPS triggers. Although not explicitly shown, the overall gain in reconstructed events is notably greater for CC interactions than for NC ones. It was also observed that events reconstructed using all triggers typically have a larger multiplicity i.e. more stations enter in the reconstruction, increasing the possibility of the said event being selected as a neutrino candidate.

\subsection{$\nu$ Selection}

\begin{figure}[ht!]
  \centering
  \includegraphics[width=0.85\textwidth]{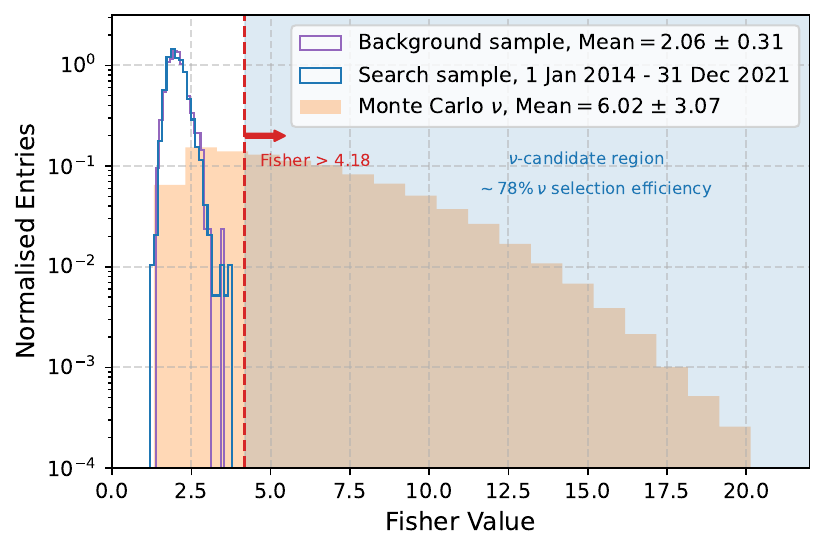}
  \caption{Distribution of the Fisher variable after the DGL event selection
for events with reconstructed zenith angle $ \theta_{\text{rec}} \in [67.5^{\circ}, 70.5^{\circ}$]. The open histograms show the background training sample (purple) and the search sample (blue) and the filled histogram (orange) depicts the simulated DGL $\nu$ events. Events above the Fisher value indicated by the vertical red dashed line would be regarded as neutrino candidates populating the blue-colored region.}
  \label{fig:Fish_67.5_70.5}
\end{figure}

Neutrino selection in the DGL range builds on the method in~\citep{auger_neutrino_limit,gap_note_2013}, with refinements to include ToTd and MoPS triggers. Events must be fully contained within the SD, have a well-reconstructed zenith angle, and at least four triggered stations. A stricter EM condition is applied: 75\% of stations closest to the core must have ToT, MoPS, or ToTd triggers to better reject background.

As in earlier work, Fisher Discriminant Analysis (FDA)~\citep{Fisher:1936et} is used with five zenith sub-ranges to incorporate shower age. However, the discriminant now uses the sum of AoP values from the four earliest triggered stations near the core — shown via simulation to improve separation. The FDA is trained on 20\% of the data selected at random from the analysis period to evaluate a detection threshold such that the expected background is fewer than one event in 20 years for the entire zenith angle range. Figure~\ref{fig:Fish_67.5_70.5} shows the discriminant distribution for data and simulated $\nu$-induced showers along with the evaluated cut in the $\theta_{\mathrm{rec}} \in (67.5^\circ, 70.5^\circ]$ sub-range. Applying this selection, a search for neutrino-induced EASs was performed in the Observatory data between 1$^\mathrm{st}$ January 2014 up to 31$^\mathrm{st}$ December 2021. No neutrino candidates were found in any of the five DGL sub-regions. 

\section{Detector exposure and limits to the diffuse flux of UHE$\nu$s in the DGL range}

\begin{figure}[ht!]
  \centering
  \subcaptionbox{Total exposure vs energy along with estimated systematic uncertainties. \label{sub:fig:Exp_total_comp}}{\includegraphics[width=.48\linewidth]{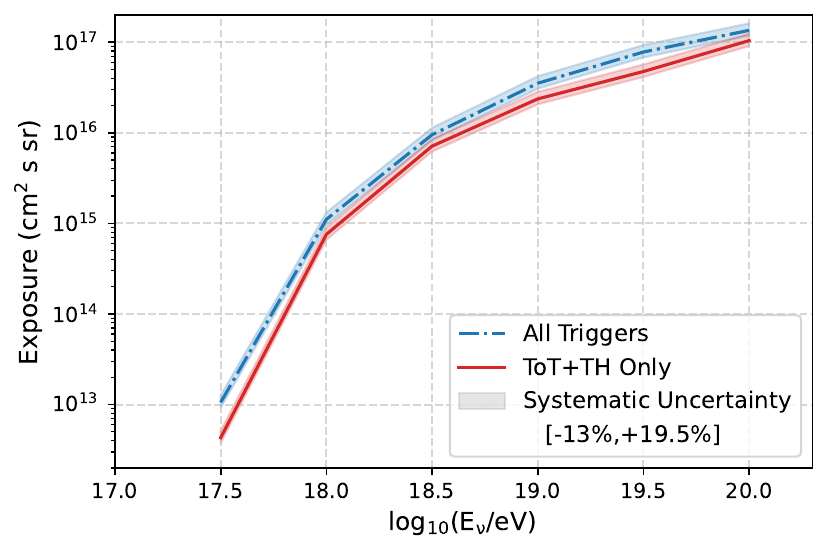}}
  \hfill
  \subcaptionbox{Average exposure per day for different neutrino energies as a function of declination. \label{sub:fig:Exp_dec_comp}}{\includegraphics[width=.48\linewidth]{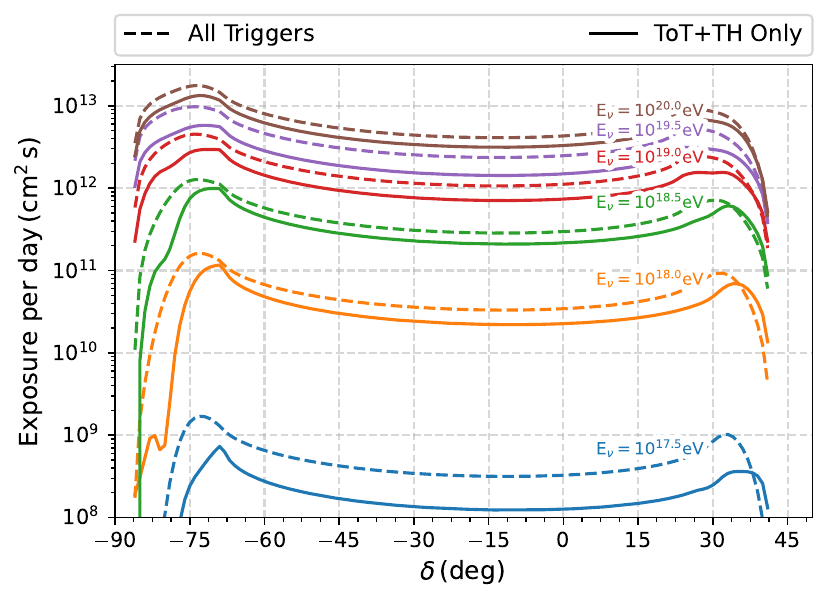}}

  \caption{Comparisons of exposure to UHE$\nu$ in the DGL angular range for the time period 1 Jan 2014- 31 Dec 2021. The dashed lines are the exposures for this analysis, and the solid lines for the analysis performed with only ToT and TH triggers for the same time period.} 
  \label{fig:Exps_comp}
\end{figure}

To use the non-observation of neutrino candidates for estimating an upper limit to the diffuse flux of UHE$\nu$s in the DGL range, the exposure is estimated using MC simulations and the triangular shape of the SD grid. A hexagon is treated as the smallest effective detection unit for neutrino events, with a Brillouin effective area $A_{\mathrm{hex}} = 1.95\,\mathrm{km}^2$. The detector efficiency, $\varepsilon^{i,c}$ for a given hexagon depends on neutrino energy $E_{\nu}$, interaction slant depth $X$, flavor \textit{i}, interaction type \textit{c} (CC or NC) and zenith angle $\theta$, and the ground impact point of the air shower.
Since the configuration of the SD array changes over time, the number of functional hexagons, $n_{\text{hex}}(t)$, recorded every minute, is used to model the evolution of the detector. The exposure $\xi^{i,c}(E_{\nu})$ is then calculated by folding in the neutrino-nucleon cross-section $\sigma^{i,c}(E_{\nu})$, the efficiency integrated over the parameter space $(X, \theta)$, and $n_{\text{hex}}(t)$, as shown in Eq.~\eqref{eq:exp},
\begin{equation}
  \xi^{i,c}(E_{\nu}) =  \frac{\sigma^{i,c}(E_{\nu})}{m_N} \,\, 2\pi \int_X \int_\theta  \,  \varepsilon^{i,c}(E_{\nu}, X, \theta) \, \sin \theta \cos \theta \, d\theta \, dX \int_{t} A_{\mathrm{hex}}\, n_{\text{hex}}(t) \, dt \,.
  \label{eq:exp}
\end{equation}
 The total integrated number of hexagons during the search period is $N_{\text{hex}} = 2.8 \times 10^{11}\,$s, an equivalent of $\sim$ 6.75 years of full-array operation. A total exposure, $\xi_{\mathrm{tot}}(E_{\nu}) = \sum_{i}\sum_{c} \xi^{i,c}(E_{\nu})$ is calculated by summing all the interaction channels and assuming a 1:1:1 flavour ratio at the earth,  
and is shown in Fig.~\ref{sub:fig:Exp_total_comp}. The overall exposure benefits significantly from the addition of the new MoPS and ToTd triggers, particularly at lower energies (up to a 5$\times$ increase) especially in the $\nu_e$ CC channel, and from the improved FDA at higher energies. The electron neutrino CC interaction dominates the total exposure ($\sim$85\%), while the NC contribution remains modest ($\sim$5\%). 
 
Assuming a power-law differential flux per unit area $A$, energy, solid angle $\Omega$, and time of the form $\phi(E_{\nu}) = \frac{d^6 N_{\nu}}{dE_{\nu}\,d\Omega\, dA \, dt} = k \cdot E_{\nu}^{-2}$, an integrated upper limit on $k$ is given by:
\begin{equation}
  \label{eq:integ_lim}
  k^{\rm DGL}_{\text{90}} = \frac{N_{\text{90}}}{\int_{E_{\nu}} E_{\nu}^{-2} \cdot \xi_{\text{tot}}(E_{\nu}) \cdot dE_{\nu}} \, ,
\end{equation}
where $N_{\text{90}}$ is the upper limit on the number of expected signal events for zero background and a 90\% confidence level. In this work, the Feldman–Cousins approach~\citep{Feldman:1997qc} extended to include systematic uncertainties~\citep{PierreAuger:2015ihf,Conrad:2002kn}, is adopted yielding $N_{\text{90}}$ = 2.39. The systematic uncertainties on exposure are $[-13\%, +19.5\%]$ taking into account the shortcomings of the neutrino simulations and their reconstruction, active hexagon counting and theoretical uncertainties associated with neutrino cross-section estimations at high energies.  

\begin{figure}[ht!]
  \centering
  \includegraphics[width=14.0cm]{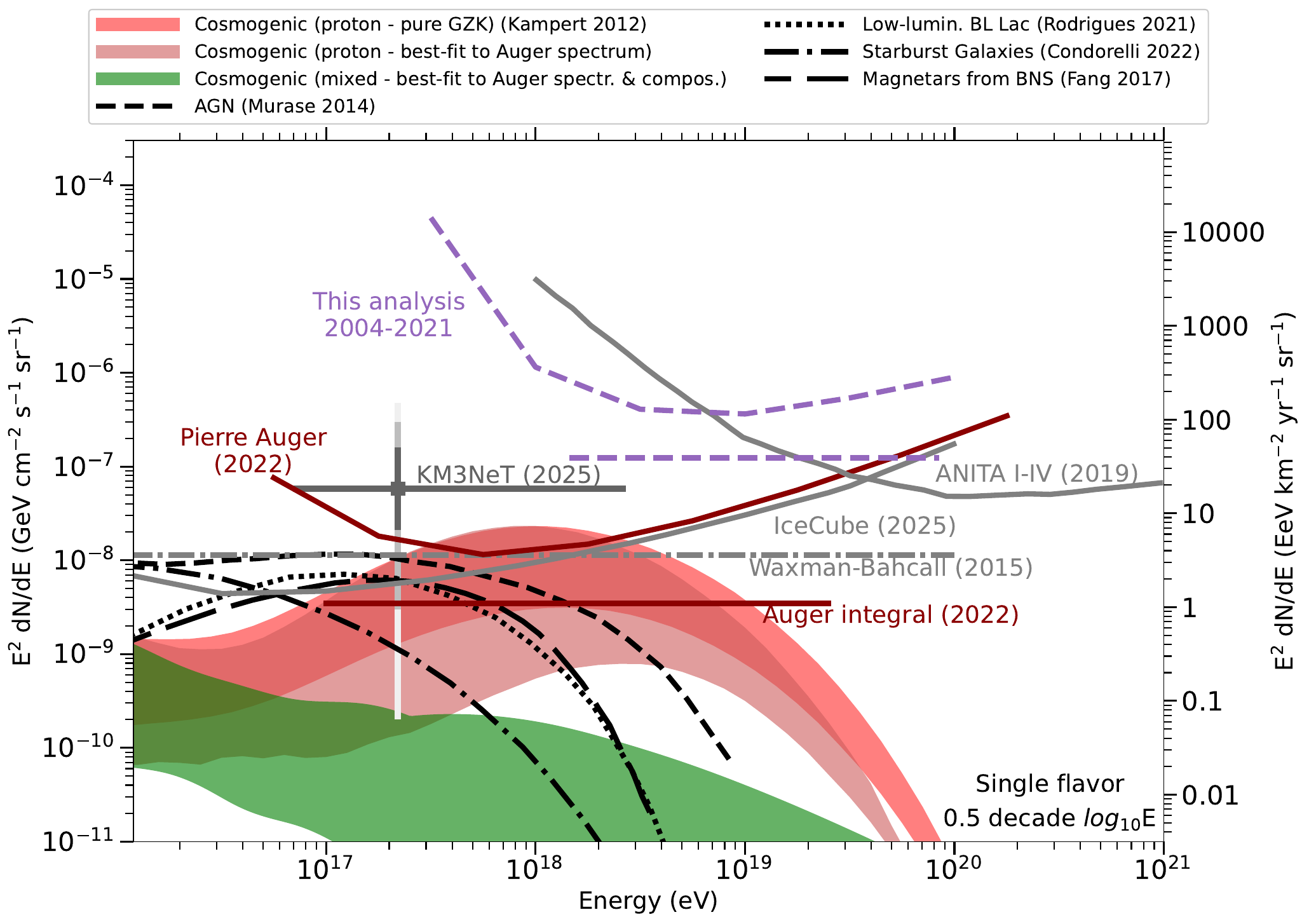}
  \caption{Comparison of the \textit{limits} (1 Jan 2004-31 Dec 2013 with \textit{ToT+TH} triggers and 1  Jan 2014-31 Dec 2021 with \textit{All} triggers) to the current upper limits on the diffuse flux of UHE neutrinos. IceCube limits from~\cite{IceCube:2025ezc} are scaled for a $E_{\nu}^{-2}$ flux assumption. The predicted fluxes from a few cosmogenic and astrophysical $\nu$ models are also shown.
  }
  \label{fig:Limit_comp_overall}
\end{figure}

The single flavour 90\% C.L. integrated limit gives $k^{\rm DGL}_{90} < 1.2 \times 10^{-7} \,\mathrm{GeV \,cm^{-2} s^{-1} sr^{-1}}$ in the DGL channel only. It applies to an energy range of $E_{\nu} \in [1.3 \times 10^{18} - 2.5 \times 10^{19.5}]$ in which $\sim$90\% of the total event rate is expected in the case of an $E^{-2}_{\nu}$ flux and is shown as a dashed purple line in Fig.~\ref{fig:Limit_comp_overall}. A \textit{differential limit} can also be calculated by integrating the denominator of Eq.~\eqref{eq:integ_lim} in bins of width $\Delta = 0.5\,$ in $\log_{10}(E_{\nu})$. The inclusion of ToTd and MoPS triggers leads to a 25\% improvement of the estimated $k^{\rm DGL}_{90}$. Although the DGL channel contributes little to the overall diffuse flux limit---still dominated by ES searches---it remains valuable for point-source studies due to its distinct field of view as discussed next.
 
\section{Search for point-like sources of UHE$\nu$s in the DGL range}

Further, a dedicated search for point-like sources of UHE neutrinos using data collected in the DGL zenith angle range with the SD array of the Pierre Auger Observatory was also performed using the new ToTd and MoPS triggers. The method relies on the time-dependent visibility of candidate sources, which depends on the source declination $\delta$ and the corresponding zenith angle $\theta(t)$, expressed as
$\cos \theta(t) = \sin \lambda \,\sin \delta + \cos \lambda \, \cos \delta \,\sin (2\pi t/T - \alpha)$,
where $\lambda$ is the latitude of the observatory and $T$ is the duration of a sidereal day. A declination-dependent exposure $\xi(E_\nu,\delta)$ shown in Fig.~\ref{sub:fig:Exp_dec_comp} was calculated by estimating the time-dependent efficiencies and replacing them in Eq.~\eqref{eq:exp}. The exposure is largest for the declinations the source is seen the longest. The exposure is assumed to be uniform in right ascension within $\pm 0.6\%$ for the time period of search as shown in~\citep{PierreAuger:2017pzq}. The non-detection of any neutrino candidates in the time period explored allowed us to set upper limits on the flux from point sources. Assuming a differential flux of the form $\phi(E_{\nu}) = k^{\text{PS}} E_{\nu}^{-2}$, the corresponding 90\% confidence level upper limit on the normalization $k^{\rm PS,DGL}_{90}$ is:

\begin{equation}
  k^{\rm PS,DGL}_{90} = 
  \frac{N_{90}}
  {\int_{E_{\text{min}}}^{E_{\text{max}}} E_{\nu}^{-2} \, \xi(E_{\nu}, \delta) \, dE_{\nu}}.
\end{equation}

\begin{figure}[ht!]
  \centering
  \includegraphics[width=0.95\textwidth]{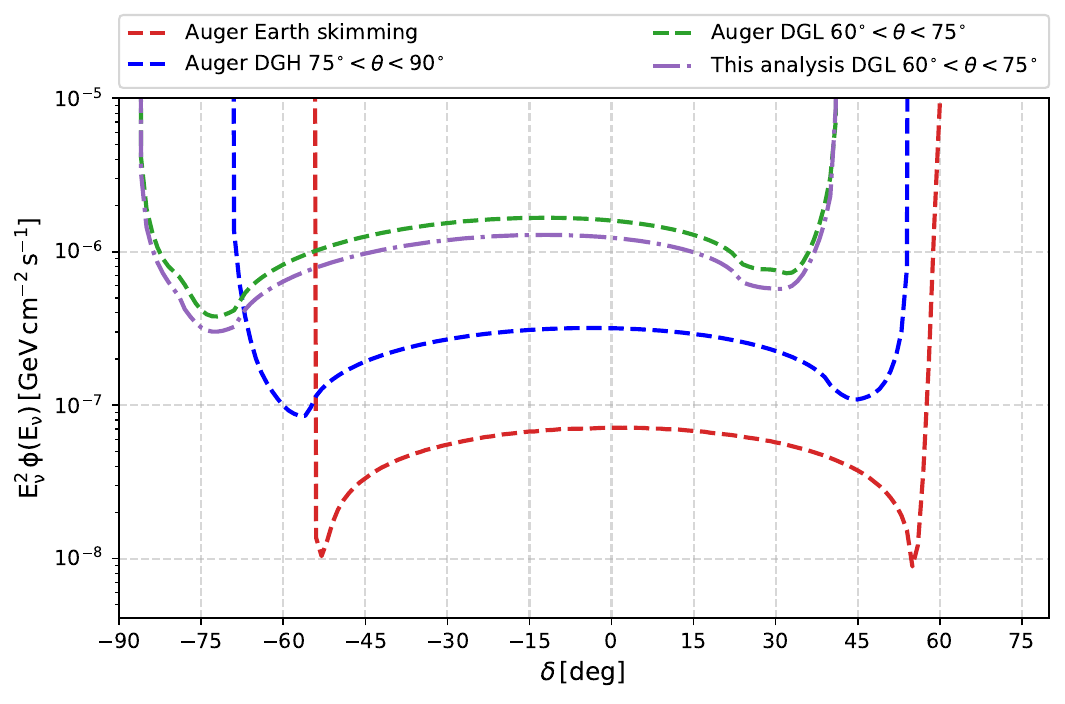}
  \caption{The upper limits  (01.01.2004–31.12.2021) at 90\% C.L. for different neutrino searches performed at the Pierre Auger Observatory of a single flavor point-like flux of UHE. The limit obtained in this analysis (purple) for the DGL channel is compared to the limits obtained for the DGH and ES channels~\citep{Aab_2019_point}. }
  \label{fig:Dec_limit_comb3}
\end{figure}

Thanks to the improved EM triggers and updated discriminant methods, this analysis achieves a 1.5-fold improvement at the most sensitive declinations compared to the DGL analysis with only ToT and TH triggers. These results are contextualized within the broader zenith-angle neutrino search program at Auger, including comparisons with limits from the DGH and ES channels in Fig.~\ref{fig:Dec_limit_comb3}.
The inclusion of ToTd and MoPS triggers increases the neutrino detection efficiency at Auger between declinations $\delta \in (\simeq -85^{\circ}, \simeq 40^{\circ})$,  
with a certain portion of the sky corresponding to $\delta\lesssim-68^\circ$ only visible in the DGL channel in comparison to other analyses.

\section{Summary and Outlook}

The Pierre Auger Observatory offers a large exposure to UHE neutrinos. The detector is continually improved to extend and enhance its detection capabilities. The analysis presented here quantifies one of these improvements by incorporating the triggers — Time-over-Threshold-deconvolved (ToTd) and Multiplicity-of-Positive-Steps (MoPS) — which were introduced to improve detection efficiency to small signals induced by the electromagnetic component of the shower. 
This analysis presents updated searches for UHE$\nu$s in the DGL zenith angle range $\theta\in[60^\circ,75^\circ]$, incorporating all triggers. A novel selection built on an earlier analysis was introduced to search for neutrino candidates in 7 years of data.   
For the diffuse flux, the enhanced selection efficiency enabled a $\sim 25\%$ more stringent 90\% C.L. upper limit, when compared to the previous selection, taking into account the DGL channel only. In point-source searches, improved detection efficiency led to increased directional exposure and stricter declination-dependent upper limits. Although the absolute gains in sensitivity obtained in this work are relatively modest, and the Pierre Auger limit is still dominated by the Earth-Skimming channel, they are crucial, especially in the $E_\nu \gtrsim 10^{18}$\,eV regime and especially in declinations $\delta\lesssim-68^\circ$ exclusively accessible in DGL. These results underline the scientific value of leveraging ToTd and MoPS triggered data and motivate further refinements to reconstruction algorithms for future UHE neutrino searches.

\providecommand{\href}[2]{#2}\begingroup\raggedright\endgroup

\clearpage

\section*{The Pierre Auger Collaboration}
\small

\begin{wrapfigure}[8]{l}{0.11\linewidth}
\vspace{-5mm}
\includegraphics[width=0.98\linewidth]{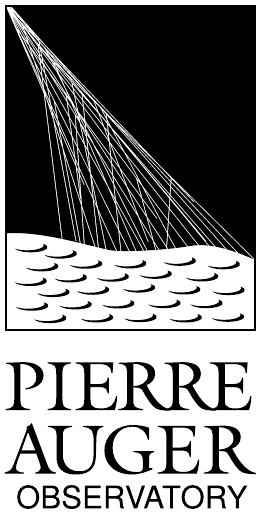}
\end{wrapfigure}

\begin{sloppypar}\noindent
A.~Abdul Halim$^{13}$,
P.~Abreu$^{70}$,
M.~Aglietta$^{53,51}$,
I.~Allekotte$^{1}$,
K.~Almeida Cheminant$^{78,77}$,
A.~Almela$^{7,12}$,
R.~Aloisio$^{44,45}$,
J.~Alvarez-Mu\~niz$^{76}$,
A.~Ambrosone$^{44}$,
J.~Ammerman Yebra$^{76}$
% ...

\end{sloppypar}

\begin{center}
\rule{0.1\columnwidth}{0.5pt}
\raisebox{-0.4ex}{\scriptsize$\bullet$}
\rule{0.1\columnwidth}{0.5pt}
\end{center}

\vspace{-1ex}
\footnotesize
\begin{description}[labelsep=0.2em,align=right,labelwidth=0.7em,labelindent=0em,leftmargin=2em,noitemsep,before={\renewcommand\makelabel[1]{##1 }}]
\item[$^{1}$] Centro At\'omico Bariloche and Instituto Balseiro (CNEA-UNCuyo-CONICET), San Carlos de Bariloche, Argentina
\item[$^{2}$] Departamento de F\'\i{}sica and Departamento de Ciencias de la Atm\'osfera y los Oc\'eanos, FCEyN, Universidad de Buenos Aires and CONICET, Buenos Aires, Argentina
\item[$^{3}$] IFLP, Universidad Nacional de La Plata and CONICET, La Plata, Argentina
\item[$^{4}$] Instituto de Astronom\'\i{}a y F\'\i{}sica del Espacio (IAFE, CONICET-UBA), Buenos Aires, Argentina
\item[$^{5}$] Instituto de F\'\i{}sica de Rosario (IFIR) -- CONICET/U.N.R.\ and Facultad de Ciencias Bioqu\'\i{}micas y Farmac\'euticas U.N.R., Rosario, Argentina
\item[$^{6}$] Instituto de Tecnolog\'\i{}as en Detecci\'on y Astropart\'\i{}culas (CNEA, CONICET, UNSAM), and Universidad Tecnol\'ogica Nacional -- Facultad Regional Mendoza (CONICET/CNEA), Mendoza, Argentina
\item[$^{7}$] Instituto de Tecnolog\'\i{}as en Detecci\'on y Astropart\'\i{}culas (CNEA, CONICET, UNSAM), Buenos Aires, Argentina
\item[$^{8}$] International Center of Advanced Studies and Instituto de Ciencias F\'\i{}sicas, ECyT-UNSAM and CONICET, Campus Miguelete -- San Mart\'\i{}n, Buenos Aires, Argentina
\item[$^{9}$] Laboratorio Atm\'osfera -- Departamento de Investigaciones en L\'aseres y sus Aplicaciones -- UNIDEF (CITEDEF-CONICET), Argentina
\item[$^{10}$] Observatorio Pierre Auger, Malarg\"ue, Argentina
\item[$^{11}$] Observatorio Pierre Auger and Comisi\'on Nacional de Energ\'\i{}a At\'omica, Malarg\"ue, Argentina
\item[$^{12}$] Universidad Tecnol\'ogica Nacional -- Facultad Regional Buenos Aires, Buenos Aires, Argentina
\item[$^{13}$] University of Adelaide, Adelaide, S.A., Australia
\item[$^{14}$] Universit\'e Libre de Bruxelles (ULB), Brussels, Belgium
\item[$^{15}$] Vrije Universiteit Brussels, Brussels, Belgium
\item[$^{16}$] Centro Brasileiro de Pesquisas Fisicas, Rio de Janeiro, RJ, Brazil
\item[$^{17}$] Centro Federal de Educa\c{c}\~ao Tecnol\'ogica Celso Suckow da Fonseca, Petropolis, Brazil
\item[$^{18}$] Instituto Federal de Educa\c{c}\~ao, Ci\^encia e Tecnologia do Rio de Janeiro (IFRJ), Brazil
\item[$^{19}$] Universidade de S\~ao Paulo, Escola de Engenharia de Lorena, Lorena, SP, Brazil
\item[$^{20}$] Universidade de S\~ao Paulo, Instituto de F\'\i{}sica de S\~ao Carlos, S\~ao Carlos, SP, Brazil
\item[$^{21}$] Universidade de S\~ao Paulo, Instituto de F\'\i{}sica, S\~ao Paulo, SP, Brazil
\item[$^{22}$] Universidade Estadual de Campinas (UNICAMP), IFGW, Campinas, SP, Brazil
\item[$^{23}$] Universidade Estadual de Feira de Santana, Feira de Santana, Brazil
\item[$^{24}$] Universidade Federal de Campina Grande, Centro de Ciencias e Tecnologia, Campina Grande, Brazil
\item[$^{25}$] Universidade Federal do ABC, Santo Andr\'e, SP, Brazil
\item[$^{26}$] Universidade Federal do Paran\'a, Setor Palotina, Palotina, Brazil
\item[$^{27}$] Universidade Federal do Rio de Janeiro, Instituto de F\'\i{}sica, Rio de Janeiro, RJ, Brazil
\item[$^{28}$] Universidad de Medell\'\i{}n, Medell\'\i{}n, Colombia
\item[$^{29}$] Universidad Industrial de Santander, Bucaramanga, Colombia
\item[$^{30}$] Charles University, Faculty of Mathematics and Physics, Institute of Particle and Nuclear Physics, Prague, Czech Republic
\item[$^{31}$] Institute of Physics of the Czech Academy of Sciences, Prague, Czech Republic
\item[$^{32}$] Palacky University, Olomouc, Czech Republic
\item[$^{33}$] CNRS/IN2P3, IJCLab, Universit\'e Paris-Saclay, Orsay, France
\item[$^{34}$] Laboratoire de Physique Nucl\'eaire et de Hautes Energies (LPNHE), Sorbonne Universit\'e, Universit\'e de Paris, CNRS-IN2P3, Paris, France
\item[$^{35}$] Univ.\ Grenoble Alpes, CNRS, Grenoble Institute of Engineering Univ.\ Grenoble Alpes, LPSC-IN2P3, 38000 Grenoble, France
\item[$^{36}$] Universit\'e Paris-Saclay, CNRS/IN2P3, IJCLab, Orsay, France
\item[$^{37}$] Bergische Universit\"at Wuppertal, Department of Physics, Wuppertal, Germany
\item[$^{38}$] Karlsruhe Institute of Technology (KIT), Institute for Experimental Particle Physics, Karlsruhe, Germany
\item[$^{39}$] Karlsruhe Institute of Technology (KIT), Institut f\"ur Prozessdatenverarbeitung und Elektronik, Karlsruhe, Germany
\item[$^{40}$] Karlsruhe Institute of Technology (KIT), Institute for Astroparticle Physics, Karlsruhe, Germany
\item[$^{41}$] RWTH Aachen University, III.\ Physikalisches Institut A, Aachen, Germany
\item[$^{42}$] Universit\"at Hamburg, II.\ Institut f\"ur Theoretische Physik, Hamburg, Germany
\item[$^{43}$] Universit\"at Siegen, Department Physik -- Experimentelle Teilchenphysik, Siegen, Germany
\item[$^{44}$] Gran Sasso Science Institute, L'Aquila, Italy
\item[$^{45}$] INFN Laboratori Nazionali del Gran Sasso, Assergi (L'Aquila), Italy
\item[$^{46}$] INFN, Sezione di Catania, Catania, Italy
\item[$^{47}$] INFN, Sezione di Lecce, Lecce, Italy
\item[$^{48}$] INFN, Sezione di Milano, Milano, Italy
\item[$^{49}$] INFN, Sezione di Napoli, Napoli, Italy
\item[$^{50}$] INFN, Sezione di Roma ``Tor Vergata'', Roma, Italy
\item[$^{51}$] INFN, Sezione di Torino, Torino, Italy
\item[$^{52}$] Istituto di Astrofisica Spaziale e Fisica Cosmica di Palermo (INAF), Palermo, Italy
\item[$^{53}$] Osservatorio Astrofisico di Torino (INAF), Torino, Italy
\item[$^{54}$] Politecnico di Milano, Dipartimento di Scienze e Tecnologie Aerospaziali , Milano, Italy
\item[$^{55}$] Universit\`a del Salento, Dipartimento di Matematica e Fisica ``E.\ De Giorgi'', Lecce, Italy
\item[$^{56}$] Universit\`a dell'Aquila, Dipartimento di Scienze Fisiche e Chimiche, L'Aquila, Italy
\item[$^{57}$] Universit\`a di Catania, Dipartimento di Fisica e Astronomia ``Ettore Majorana``, Catania, Italy
\item[$^{58}$] Universit\`a di Milano, Dipartimento di Fisica, Milano, Italy
\item[$^{59}$] Universit\`a di Napoli ``Federico II'', Dipartimento di Fisica ``Ettore Pancini'', Napoli, Italy
\item[$^{60}$] Universit\`a di Palermo, Dipartimento di Fisica e Chimica ''E.\ Segr\`e'', Palermo, Italy
\item[$^{61}$] Universit\`a di Roma ``Tor Vergata'', Dipartimento di Fisica, Roma, Italy
\item[$^{62}$] Universit\`a Torino, Dipartimento di Fisica, Torino, Italy
\item[$^{63}$] Benem\'erita Universidad Aut\'onoma de Puebla, Puebla, M\'exico
\item[$^{64}$] Unidad Profesional Interdisciplinaria en Ingenier\'\i{}a y Tecnolog\'\i{}as Avanzadas del Instituto Polit\'ecnico Nacional (UPIITA-IPN), M\'exico, D.F., M\'exico
\item[$^{65}$] Universidad Aut\'onoma de Chiapas, Tuxtla Guti\'errez, Chiapas, M\'exico
\item[$^{66}$] Universidad Michoacana de San Nicol\'as de Hidalgo, Morelia, Michoac\'an, M\'exico
\item[$^{67}$] Universidad Nacional Aut\'onoma de M\'exico, M\'exico, D.F., M\'exico
\item[$^{68}$] Institute of Nuclear Physics PAN, Krakow, Poland
\item[$^{69}$] University of \L{}\'od\'z, Faculty of High-Energy Astrophysics,\L{}\'od\'z, Poland
\item[$^{70}$] Laborat\'orio de Instrumenta\c{c}\~ao e F\'\i{}sica Experimental de Part\'\i{}culas -- LIP and Instituto Superior T\'ecnico -- IST, Universidade de Lisboa -- UL, Lisboa, Portugal
\item[$^{71}$] ``Horia Hulubei'' National Institute for Physics and Nuclear Engineering, Bucharest-Magurele, Romania
\item[$^{72}$] Institute of Space Science, Bucharest-Magurele, Romania
\item[$^{73}$] Center for Astrophysics and Cosmology (CAC), University of Nova Gorica, Nova Gorica, Slovenia
\item[$^{74}$] Experimental Particle Physics Department, J.\ Stefan Institute, Ljubljana, Slovenia
\item[$^{75}$] Universidad de Granada and C.A.F.P.E., Granada, Spain
\item[$^{76}$] Instituto Galego de F\'\i{}sica de Altas Enerx\'\i{}as (IGFAE), Universidade de Santiago de Compostela, Santiago de Compostela, Spain
\item[$^{77}$] IMAPP, Radboud University Nijmegen, Nijmegen, The Netherlands
\item[$^{78}$] Nationaal Instituut voor Kernfysica en Hoge Energie Fysica (NIKHEF), Science Park, Amsterdam, The Netherlands
\item[$^{79}$] Stichting Astronomisch Onderzoek in Nederland (ASTRON), Dwingeloo, The Netherlands
\item[$^{80}$] Universiteit van Amsterdam, Faculty of Science, Amsterdam, The Netherlands
\item[$^{81}$] Case Western Reserve University, Cleveland, OH, USA
\item[$^{82}$] Colorado School of Mines, Golden, CO, USA
\item[$^{83}$] Department of Physics and Astronomy, Lehman College, City University of New York, Bronx, NY, USA
\item[$^{84}$] Michigan Technological University, Houghton, MI, USA
\item[$^{85}$] New York University, New York, NY, USA
\item[$^{86}$] University of Chicago, Enrico Fermi Institute, Chicago, IL, USA
\item[$^{87}$] University of Delaware, Department of Physics and Astronomy, Bartol Research Institute, Newark, DE, USA
\item[] -----
\item[$^{a}$] Max-Planck-Institut f\"ur Radioastronomie, Bonn, Germany
\item[$^{b}$] also at Kapteyn Institute, University of Groningen, Groningen, The Netherlands
\item[$^{c}$] School of Physics and Astronomy, University of Leeds, Leeds, United Kingdom
\item[$^{d}$] Fermi National Accelerator Laboratory, Fermilab, Batavia, IL, USA
\item[$^{e}$] Pennsylvania State University, University Park, PA, USA
\item[$^{f}$] Colorado State University, Fort Collins, CO, USA
\item[$^{g}$] Louisiana State University, Baton Rouge, LA, USA
\item[$^{h}$] now at Graduate School of Science, Osaka Metropolitan University, Osaka, Japan
\item[$^{i}$] Institut universitaire de France (IUF), France
\item[$^{j}$] now at Technische Universit\"at Dortmund and Ruhr-Universit\"at Bochum, Dortmund and Bochum, Germany
\end{description}

\vspace{-1ex}
\footnotesize
\section*{Acknowledgments}

\begin{sloppypar}
The successful installation, commissioning, and operation of the Pierre
Auger Observatory would not have been possible without the strong
commitment and effort from the technical and administrative staff in
Malarg\"ue. We are very grateful to the following agencies and
organizations for financial support:
\end{sloppypar}

\begin{sloppypar}
Argentina -- Comisi\'on Nacional de Energ\'\i{}a At\'omica; Agencia Nacional de
Promoci\'on Cient\'\i{}fica y Tecnol\'ogica (ANPCyT); Consejo Nacional de
Investigaciones Cient\'\i{}ficas y T\'ecnicas (CONICET); Gobierno de la
Provincia de Mendoza; Municipalidad de Malarg\"ue; NDM Holdings and Valle
Las Le\~nas; in gratitude for their continuing cooperation over land
access; Australia -- the Australian Research Council; Belgium -- Fonds
de la Recherche Scientifique (FNRS); Research Foundation Flanders (FWO),
Marie Curie Action of the European Union Grant No.~101107047; Brazil --
Conselho Nacional de Desenvolvimento Cient\'\i{}fico e Tecnol\'ogico (CNPq);
Financiadora de Estudos e Projetos (FINEP); Funda\c{c}\~ao de Amparo \`a
Pesquisa do Estado de Rio de Janeiro (FAPERJ); S\~ao Paulo Research
Foundation (FAPESP) Grants No.~2019/10151-2, No.~2010/07359-6 and
No.~1999/05404-3; Minist\'erio da Ci\^encia, Tecnologia, Inova\c{c}\~oes e
Comunica\c{c}\~oes (MCTIC); Czech Republic -- GACR 24-13049S, CAS LQ100102401,
MEYS LM2023032, CZ.02.1.01/0.0/0.0/16{\textunderscore}013/0001402,
CZ.02.1.01/0.0/0.0/18{\textunderscore}046/0016010 and
CZ.02.1.01/0.0/0.0/17{\textunderscore}049/0008422 and CZ.02.01.01/00/22{\textunderscore}008/0004632;
France -- Centre de Calcul IN2P3/CNRS; Centre National de la Recherche
Scientifique (CNRS); Conseil R\'egional Ile-de-France; D\'epartement
Physique Nucl\'eaire et Corpusculaire (PNC-IN2P3/CNRS); D\'epartement
Sciences de l'Univers (SDU-INSU/CNRS); Institut Lagrange de Paris (ILP)
Grant No.~LABEX ANR-10-LABX-63 within the Investissements d'Avenir
Programme Grant No.~ANR-11-IDEX-0004-02; Germany -- Bundesministerium
f\"ur Bildung und Forschung (BMBF); Deutsche Forschungsgemeinschaft (DFG);
Finanzministerium Baden-W\"urttemberg; Helmholtz Alliance for
Astroparticle Physics (HAP); Helmholtz-Gemeinschaft Deutscher
Forschungszentren (HGF); Ministerium f\"ur Kultur und Wissenschaft des
Landes Nordrhein-Westfalen; Ministerium f\"ur Wissenschaft, Forschung und
Kunst des Landes Baden-W\"urttemberg; Italy -- Istituto Nazionale di
Fisica Nucleare (INFN); Istituto Nazionale di Astrofisica (INAF);
Ministero dell'Universit\`a e della Ricerca (MUR); CETEMPS Center of
Excellence; Ministero degli Affari Esteri (MAE), ICSC Centro Nazionale
di Ricerca in High Performance Computing, Big Data and Quantum
Computing, funded by European Union NextGenerationEU, reference code
CN{\textunderscore}00000013; M\'exico -- Consejo Nacional de Ciencia y Tecnolog\'\i{}a
(CONACYT) No.~167733; Universidad Nacional Aut\'onoma de M\'exico (UNAM);
PAPIIT DGAPA-UNAM; The Netherlands -- Ministry of Education, Culture and
Science; Netherlands Organisation for Scientific Research (NWO); Dutch
national e-infrastructure with the support of SURF Cooperative; Poland
-- Ministry of Education and Science, grants No.~DIR/WK/2018/11 and
2022/WK/12; National Science Centre, grants No.~2016/22/M/ST9/00198,
2016/23/B/ST9/01635, 2020/39/B/ST9/01398, and 2022/45/B/ST9/02163;
Portugal -- Portuguese national funds and FEDER funds within Programa
Operacional Factores de Competitividade through Funda\c{c}\~ao para a Ci\^encia
e a Tecnologia (COMPETE); Romania -- Ministry of Research, Innovation
and Digitization, CNCS-UEFISCDI, contract no.~30N/2023 under Romanian
National Core Program LAPLAS VII, grant no.~PN 23 21 01 02 and project
number PN-III-P1-1.1-TE-2021-0924/TE57/2022, within PNCDI III; Slovenia
-- Slovenian Research Agency, grants P1-0031, P1-0385, I0-0033, N1-0111;
Spain -- Ministerio de Ciencia e Innovaci\'on/Agencia Estatal de
Investigaci\'on (PID2019-105544GB-I00, PID2022-140510NB-I00 and
RYC2019-027017-I), Xunta de Galicia (CIGUS Network of Research Centers,
Consolidaci\'on 2021 GRC GI-2033, ED431C-2021/22 and ED431F-2022/15),
Junta de Andaluc\'\i{}a (SOMM17/6104/UGR and P18-FR-4314), and the European
Union (Marie Sklodowska-Curie 101065027 and ERDF); USA -- Department of
Energy, Contracts No.~DE-AC02-07CH11359, No.~DE-FR02-04ER41300,
No.~DE-FG02-99ER41107 and No.~DE-SC0011689; National Science Foundation,
Grant No.~0450696, and NSF-2013199; The Grainger Foundation; Marie
Curie-IRSES/EPLANET; European Particle Physics Latin American Network;
and UNESCO.
\end{sloppypar}

\end{document}